\documentclass[10pt,aps,prl,twocolumn,superscriptaddress]{revtex4}
\usepackage{hyperref}
\usepackage{graphicx}
\usepackage{amsmath,amssymb}
\usepackage{subfigure}
\usepackage{url}
\usepackage{flushend}

\graphicspath{{Images/}}

\begin{document}

\title{
Phase-sensitive narrowband heterodyne holography
}

\author{Francois Bruno}

\affiliation{
Institut Langevin. Fondation Pierre-Gilles de Gennes. Centre National de la Recherche Scientifique (CNRS) UMR 7587, Institut National de la Sant\'e et de la Recherche M\'edicale (INSERM) U 979, Universit\'e Pierre et Marie Curie (UPMC), Universit\'e Paris 7. Ecole Sup\'erieure de Physique et de Chimie Industrielles (ESPCI) - 1 rue Jussieu. 75005 Paris. France
}

\author{Jean-Baptiste Laudereau}

\affiliation{
Institut Langevin. Fondation Pierre-Gilles de Gennes. Centre National de la Recherche Scientifique (CNRS) UMR 7587, Institut National de la Sant\'e et de la Recherche M\'edicale (INSERM) U 979, Universit\'e Pierre et Marie Curie (UPMC), Universit\'e Paris 7. Ecole Sup\'erieure de Physique et de Chimie Industrielles (ESPCI) - 1 rue Jussieu. 75005 Paris. France
}

\author{Max Lesaffre}

\affiliation{
Institut Langevin. Fondation Pierre-Gilles de Gennes. Centre National de la Recherche Scientifique (CNRS) UMR 7587, Institut National de la Sant\'e et de la Recherche M\'edicale (INSERM) U 979, Universit\'e Pierre et Marie Curie (UPMC), Universit\'e Paris 7. Ecole Sup\'erieure de Physique et de Chimie Industrielles (ESPCI) - 1 rue Jussieu. 75005 Paris. France
}

\author{Nicolas Verrier}

\affiliation{
Institut Langevin. Fondation Pierre-Gilles de Gennes. Centre National de la Recherche Scientifique (CNRS) UMR 7587, Institut National de la Sant\'e et de la Recherche M\'edicale (INSERM) U 979, Universit\'e Pierre et Marie Curie (UPMC), Universit\'e Paris 7. Ecole Sup\'erieure de Physique et de Chimie Industrielles (ESPCI) - 1 rue Jussieu. 75005 Paris. France
}

\author{Michael Atlan}

\affiliation{
Institut Langevin. Fondation Pierre-Gilles de Gennes. Centre National de la Recherche Scientifique (CNRS) UMR 7587, Institut National de la Sant\'e et de la Recherche M\'edicale (INSERM) U 979, Universit\'e Pierre et Marie Curie (UPMC), Universit\'e Paris 7. Ecole Sup\'erieure de Physique et de Chimie Industrielles (ESPCI) - 1 rue Jussieu. 75005 Paris. France
}

\date{\today}

\begin{abstract}
We report on amplitude and phase imaging of out-of-plane sinusoidal surface vibration at nanometer scales with a heterodyne holographic interferometer. The originality of the proposed method is to make use of a multiplexed local oscillator to address several optical sidebands into the temporal bandwidth of a sensor array. This process is called coherent frequency-division multiplexing. It enables simultaneous recording and pixel-to-pixel division of sideband holograms, which permits quantitative wide-field mapping of optical phase modulation depths. Additionally, a linear frequency chirp ensures the retrieval of the local mechanical phase shift of the vibration with respect to the excitation signal. The proposed approach is validated by quantitative motion characterization of the lamellophone of a musical box, behaving as a group of harmonic oscillators, under weak sinusoidal excitation. Images of the vibration amplitude versus excitation frequency show the resonance of the nanometric flexural response of one individual cantilever, at which a phase hop is measured.
\end{abstract}

\maketitle

Imaging nanometric optical path length modulations is of interest in non-destructive testing of micro electro-mechanical systems~\cite{Bosseboeuf2003, Rembe2006, Kokkonen2008, BramhavarPouet2009}. Among the approaches used for  single-point vibration spectra measurements, laser Doppler schemes are commonly found; they enable non-contact out-of-plane vibration monitoring~\cite{WhitmanKorpel1969, DeLaRue1972, Stegeman1976}. In these measurement schemes, the resolution of the minimum measurable vibration amplitude is limited by the detection sensitivity~\cite{WhitmanKorpel1969, Monchalin1985, WagnerSpicer1987, RoyerDieulesaint1989, Rembe2006, Kokkonen2008}. These methods exhibit high reliability for wideband~\cite{RoyerDieulesaint1986, RoyerDieulesaint1989, JiaBoumiz1993, RoyerKmetik1992} and narrowband~\cite{DeLaRue1972, Monchalin1985, BramhavarPouet2009} measurements. Narrowband measurements facilitate high frequency monitoring without the need for specialized electronics~\cite{Kokkonen2008, BramhavarPouet2009}, while wideband measurements allow for transient signal analysis~\cite{RoyerDieulesaint1986}. Nevertheless, imaging is a tedious and time-consuming process because laser probe scanning is required. To circumvent  scanning issues, wide-field heterodyne optical detection schemes with dedicated sensor arrays were introduced recently~\cite{Kimachi2010, PatelAchamfuoYeboah2011}. Another approach involves time-averaged holographic interferometry, which permits high sensitivity measurements~\cite{Powell1965, Aleksoff1969, Stetson1970, Levitt1976, UedaMiida1976, PicartLeval2003}. This method enables narrowband measurements of the optical pathlength modulation depth when the exposure time of the recorded interferogram is much longer than the vibration period~\cite{Aleksoff1969, PsotaLedl2012, VerrierAtlan2013}. With this approach, the optical phase retardation of the recorded scattered field beating against a reference optical field can also be exploited for time-resolved (wideband) determination of the object motion~\cite{Pedrini2006, PerezLopez2006}. Nevertheless, typical vibration frequencies of interest in non-destructive monitoring are still much higher than camera frame rates, and hence the time-averaging condition holds. In this regime, the phase shift with respect to the excitation signal is usually retrieved from phase-stepped stroboscopic schemes~\cite{Lokberg1976, Petitgrand2001, Bosseboeuf2003, LevalPicart2005, VerrierGrossAtlan2013}. Quantitative imaging of optical path length modulation and phase retardation in stationary regimes, at the time scale of the frame exposure, may thus find applications.\\

\begin{figure}[]
  \centering
\includegraphics[width = 8 cm]{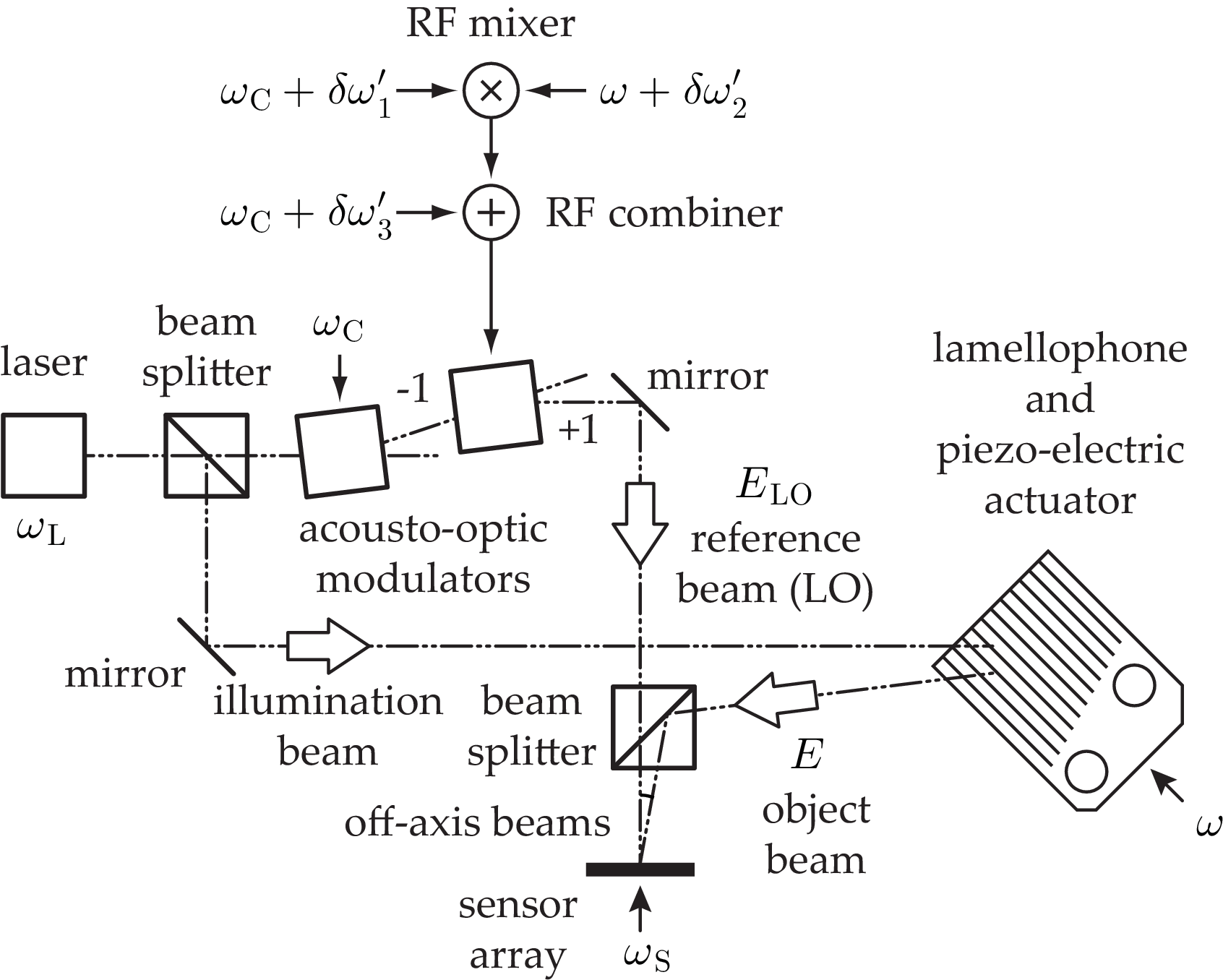}
\caption{Experimental set-up.}
\label{fig_setup}
\end{figure}

In this letter, we report on a heterodyne holographic arrangement used for amplitude and phase imaging of optical path length modulation in time-averaged recording conditions, without strobe light. The proposed arrangement consists of a Mach-Zehnder holographic interferometer in an off-axis and frequency-shifting configuration~\cite{AtlanGross2007JOSAA, VerrierAtlan2013}, with a multiplexed local oscillator (Fig.~\ref{fig_setup}). Frequency-tunable narrowband detection is achieved in time-averaged heterodyne detection conditions. Time-averaged holograms of a phase-modulated optical field are recorded with an exposure time much longer than the modulation period. Optical heterodyning is performed with a frequency-shifted optical local oscillator, a frequency-conversion process aimed at shifting a given radiofrequency (RF) optical sideband in the sensor bandwidth. Optical local oscillator multiplexing is realized by the coherent addition of RF signals driving an acousto-optic modulator in the reference beam channel. Coherent frequency-division multiplexing is a technique by which the total available bandwidth is divided into non-overlapping frequency sub-bands carrying separate signals. Addressing simultaneously distinct domains of the spectral bandwidth of holograms was performed with success for angular~\cite{PaturzoMemmolo2009},  wavelength~\cite{Kiire2012}, space-division~\cite{Tahara2012}, and sideband~\cite{VerrierAtlan2013} multiplexing schemes. Coherent frequency-division multiplexing in heterodyne holography offers the opportunity to measure complex-valued maps of the RF spectral components of the optical radiation field with accuracy and sensitivity, as long as they can be isolated in sub-bands of the sensor's temporal bandwidth. In the reported experiment, high-quality factor flexural responses at audio frequencies of a lamellophone (the metallic comb of cantilevers from a musical box) excited in a sympathetic resonance regime are sought. Sympathetic vibration is a harmonic phenomenon wherein a vibratory body responds to external excitation in the vicinity of a resonance frequency.\\

The optical setup used to perform non-destructive testing of the cantilevers' out-of-plane flexural responses at nanometric scales is sketched in Fig.~\ref{fig_setup}. It is designed to monitor quantitatively steady-state vibration amplitudes and to detect mechanical phase jumps with respect to the excitation signal, by interferometric detection of an optical field scattered by the lamellophone, referred to as the object field, beating against a reference optical field. The out-of-plane motion of the cantilevers provokes an optical path length modulation of the backscattered object field. Hence the temporal part of the object field $E$ undergoes a sinusoidal phase modulation $\phi(t) = 4\pi z(t) / \lambda$. The equation of the out-of-plane motion $z(t)$ of a cantilever, modeled as a damped harmonic oscillator of spring constant $k$, mass $m$, damping coefficient $c$, driven sinusoidally by the force $F(t) = F_0 \sin (\omega t)$ provided by a piezo-electric actuator is
\begin{equation} 
m \frac{ \partial ^2 z}{ \partial t^2 } + c \frac{\partial z}{\partial t} + k z = F(t)
\label{eq_HarmonicOscillator}
\end{equation}
We seek a steady-state solution at the excitation frequency with an induced phase change of $\psi$ with respect to the excitation $F(t)$, as a result of viscous damping
\begin{equation} 
z(t) = z_0 \sin \left( \omega t + \psi \right) 
\label{eq_z_SolutionForm}
\end{equation}
Its amplitude, $z_0$, is proportional to the driving force
\begin{equation} 
    z_0 = \frac{F_0}{m \omega} \left[ \left(2\omega_0\zeta\right)^2 + \frac{1}{\omega^2} \left(\omega_0^2 - \omega^2\right)^2 \right]^{-1/2} 
\label{eq_z0}
\end{equation}
Two intermediate variables were introduced : the angular resonance frequency $\omega_0$ of the undamped oscillator, and the dimensionless damping ratio $\zeta$. They satisfy the relationships $k = m \omega_0^2$ and $c = 2 m \omega_0 \zeta$. The phase shift of the sinusoidal response with respect to the driving force is
\begin{equation} 
    \psi = \arctan\left(\frac{2\omega \omega_0\zeta}{\omega^2-\omega_0^2}\right)
\label{eq_psi}
\end{equation}
The resonance frequencies $\omega_{\rm r}= \omega_0 \sqrt{1 - 2 \zeta^2}$ of each cantilever are almost equal to the resonant frequencies $\omega_0$ of the undamped system. The monochromatic optical field of amplitude ${\cal E}$, oscillating at the angular frequency $\omega_{\rm{L}}$, backscattered by the object in vibration can be written
\begin{equation}
E = {\cal E} \exp \left[ i \omega_{\rm{L}}t + i \phi (t) \right]
\end{equation}
The phase-modulated field can be decomposed on a basis of Bessel functions of the first kind $J_n(\phi_0)$, via the Jacobi-Anger identity
\begin{equation}
E = {\cal E} \exp(i \omega_{\rm{L}}t ) \sum_{n} J_n \left(\phi_0\right) \exp \left[ i n \left(  \omega t + \psi \right) \right]
\label{eq_Jacobi}
\end{equation}
where the quantity 
\begin{equation}
{\cal E}_n = {\cal E} J_n \left(\phi_0\right) \exp(i n \psi)
\label{eq_ComplexSidebandsWeights}
\end{equation}
is the complex weight of the optical sideband of order $n$, and $\phi_0 = 4\pi z_0/\lambda$ is the modulation depth of the optical phase of the radiation backscattered by the cantilever. The local amplitude $z_0$ of the out-of-plane vibration and phase retardation $\psi$ with respect to the sinusoidal driving force can be derived from the complex weight of two optical sidebands with Eq.~\ref{eq_ComplexSidebandsWeights} and the first-order Taylor developments of $J_0(\phi_0)$ and $J_1(\phi_0)$, near $\phi_0 = 0$, for $z_0 \ll \lambda$
\begin{equation}
\frac{{\cal E}_1}{{\cal E}_{0}}  \approx \frac{2\pi}{\lambda} 
\label{eq_z_psi} z_0 \exp(i \psi)
\end{equation}
\begin{figure}[]
\centering
\includegraphics[width = 8 cm]{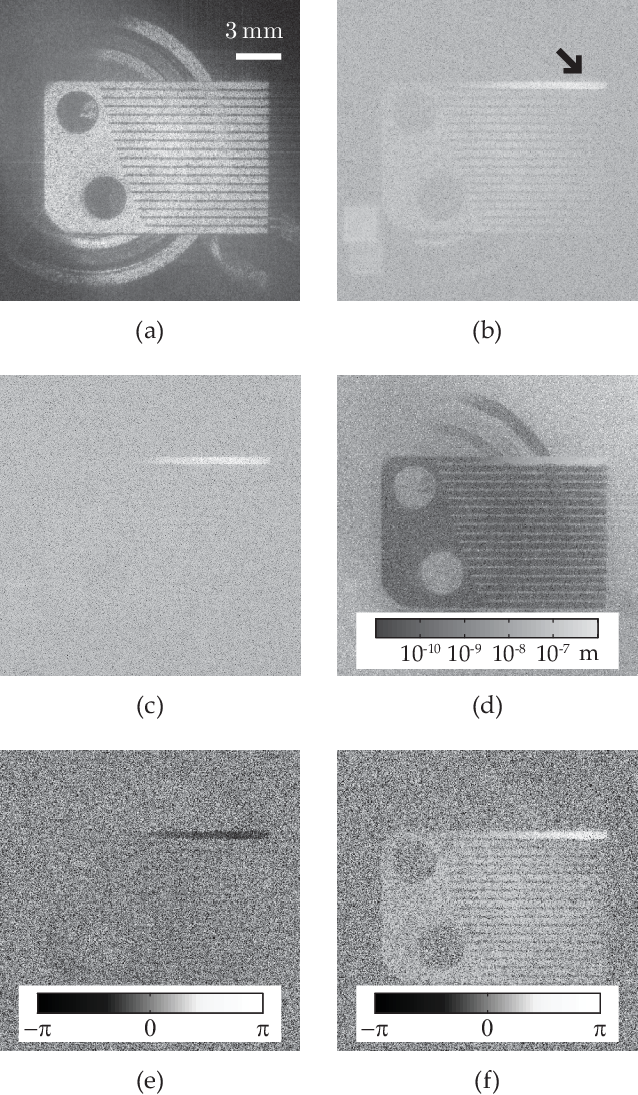}
\caption{Magnitude of a statically-scattered light hologram $|\tilde{H}_0|$ (a). Magnitude of sideband holograms $|\tilde{H}_{-1}|$ (b) and $|\tilde{H}_1|$ (c). Amplitude map of the flexural response $z_0$ of the first cantilever for the first resonance at $\omega/(2\pi) = 541 \, \rm Hz$ (d). Phase images $\psi$ calculated from 128 raw interferograms around 540.9 Hz (e) and 541.8 Hz (f), in the neighborhood of the first resonance. Movies of the amplitude and phase maps of the out-of-plane vibration are reported in media 1 and 2.}
\label{fig_AmplitudeAndPhaseImages}
\end{figure}
The holographic interferometer is used for the detection of RF phase modulation sidebands of an object field $E$ in reflective geometry, beating against a LO field $E_{\rm LO}$, as reported in~\cite{VerrierAtlan2013}. Its purpose is to enable a robust quantitative measurement of the nanometric modulation amplitudes of cantilever resonances, as well as the associated phase shift with respect to the sinusoidal excitation signal. To do so, the adopted strategy is to separate the sensor bandwidth $[ -\omega_{\rm S}/2 ,\omega_{\rm S}/2[$  into three non-overlapping frequency sub-bands, centered at $\delta \omega _1$, $\delta \omega _2$, and $\delta \omega _3$, each of which is used to carry a separate optical modulation sideband. The LO signal is composed of the sum of three phase-locked RF signals, to yield an optical LO field of the form
\begin{equation}
E_{\rm{LO}} = {\cal E}_{\rm{LO}} \exp( i\omega_{\rm{L}}t)
{\sum_{n=1}^{3}}\alpha_n \exp(i \Delta \omega_n t)
\label{eq_E_LO}
\end{equation}
where $\Delta \omega_n = (n-2)\omega - \delta \omega_n$ is the frequency shift and $ \alpha_n {\cal E}_{\rm{LO}}$ is the complex weight of the LO component of rank $n=1, 2, 3$. The positive parameters $\alpha_n$ are the normalized weights of each LO component; in this experiment $\alpha_1 = \alpha_2 = \alpha_3 = 1/3$. The LO components are tuned around the optical modulation bands of order $-1$, $0$, and $+1$. Precisely, the LO component of rank $n$ beats against the optical modulation band of rank $n-2$ at the apparent frequency $\delta \omega_n$, which is set within the camera bandwidth. The sensor array used for detection is an EMCCD camera, (Andor IXON 885+, frame rate $\omega_{\rm S} /(2\pi) = 20 \, \rm Hz$). The main optical radiation field is provided by a doubled Nd:YAG laser (Oxxius SLIM 532, power 100 mW, wavelength $\lambda = 532$ nm, optical pulsation $\omega_{\rm L}$). Image rendering from the raw interferograms ${\cal I}(t) = \left| E(t) + E_{\rm{LO}}(t) \right|^2 $ impinging on the sensor array involves a diffraction calculation performed with a numerical Fresnel transform. Its implementation in off-axis holographic conditions~\cite{Cuche2000} yields complex-valued time-averaged~\cite{Powell1965, Aleksoff1969, Stetson1970, Levitt1976} holograms $I(t)$ from which the off-axis region 
\begin{equation}
H(t) = E(t) E_{\rm{LO}}^*(t)
\end{equation}
is the support of the object-reference fields cross-term. The demodulation process (Eq.~\ref{eq_FFT_I_Itilde}) from $N$ of consecutive recordings of finite exposure times acts as a band-pass filter of width $\omega_{\rm S}/N$ in the temporal frequency domain. Only the contributions within the sensor temporal bandwidth, between the Nyquist frequencies $\pm \omega_{\rm S} / 2$, have to be taken into account in $H(t)$, which can hence be expressed as
\begin{figure}[]
\centering
\includegraphics[width = 8 cm]{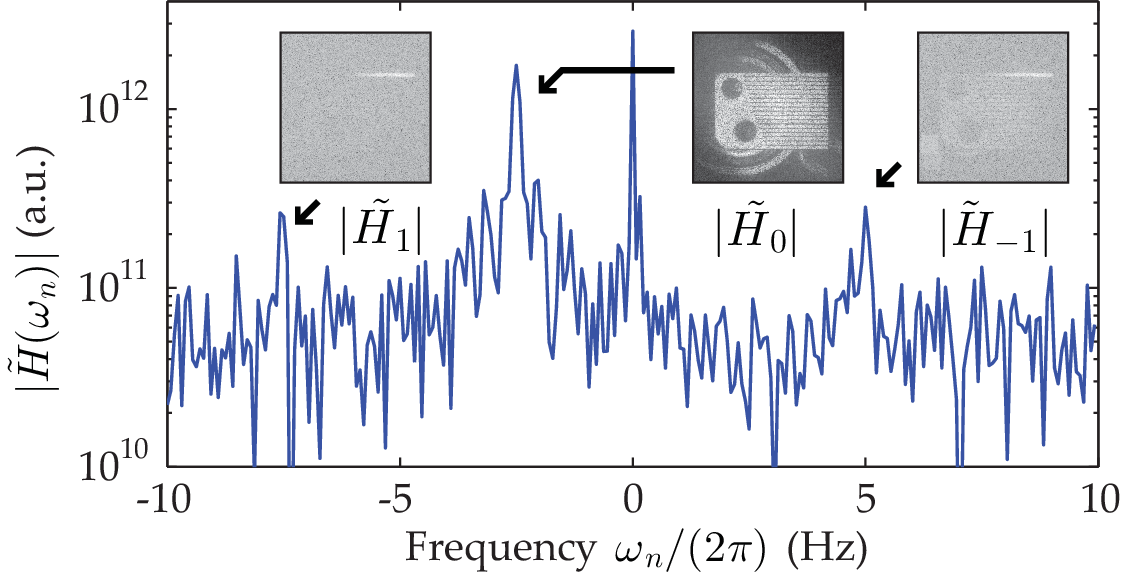}
\caption{Magnitude of the FFT-spectrum (Eq.~\ref{eq_FFT_I_Itilde} and Eq.~\ref{eq_Hm}), where three modulation sidebands are addressed.}
\label{fig_CCDSpectrum}
\end{figure}
\begin{equation}\label{eq_interferogram_in_CCD_BW}
H(t) = \sum_{n=1}^{3} {\cal E}_{n-2} \, {\cal E}^{*}_{\rm{LO}} \alpha_n {\exp}{\left( i \delta \omega_{n}t\right)}
\end{equation}
where the quantities $\delta \omega_n$ satisfy $-\omega_{\rm S} / 2 <  \delta \omega_n <\omega_{\rm S} / 2$. The FFT-spectrum of $N$ consecutive off-axis holograms
\begin{equation}
\tilde{H}\left(\omega_n\right) = \textstyle{\sum_{p=1}^N} H(2 \pi p / \omega_{\rm S}) \exp{\left(-2inp\pi/N\right)}
\label{eq_FFT_I_Itilde}
\end{equation}
exhibits three peaks at the frequencies $\delta \omega_1$, $\delta \omega_2$, and $\delta \omega_3$ (Fig.~\ref{fig_CCDSpectrum}). Their complex values yield holograms $\tilde{H}_m$, of rank $m = -1, 0, 1$, whose magnitude is proportional the modulation depth at the frequency $\delta \omega_{m+2}$
\begin{equation}\label{eq_Hm}
\tilde{H}_m = \tilde{H}\left(\delta \omega_{m+2}\right) = A \, {\cal E}_{m} 
\end{equation}
where $A \propto {\cal E}^{*}_{\rm{LO}} \alpha_{m}$ is a complex constant. Thus, $\tilde{H}_m$ is a measure of the complex weight ${\cal E}_m$. For small-amplitude vibrations, a quantitative map of the local out-of-plane vibration amplitude $z_0$ can be assessed from the ratio of the first-order sideband $\tilde{H}_{1}$ and the non-shifted light component $\tilde{H}_0$
\begin{equation}
z_0 \approx \frac{\lambda}{2\pi} \left| {\tilde{H}_1}/{\tilde{H}_{0}} \right|
\label{eq_MeasuredAmplitude}
\end{equation}
In the same manner, a map of the local mechanical phase retardation $\psi$ with respect to the excitation signal can be assessed from $\tilde{H}_{1}$, and $\tilde{H}_0$
\begin{equation}
\psi - \psi_0 = {\rm Arg} \left( {\tilde{H}_1}/{\tilde{H}_{0}} \right)
\label{eq_MeasuredPsi}
\end{equation}
where $\psi_0(t) = \delta \omega_3 t - \delta \omega_2 t$. Relationships \ref{eq_MeasuredAmplitude} and \ref{eq_MeasuredPsi} are derived from Eq.~\ref{eq_z_psi}, and Eq.~\ref{eq_Hm}. One can also make use of the third recorded band $\tilde{H}_{-1}$ and the property $J_{-1}(\phi_0) = - J_{1}(\phi_0)$ to derive $z_0 \approx \lambda/(2\pi) \left| {\tilde{H}_{-1}}/{\tilde{H}_{0}} \right|$.\\

\begin{figure}[]
\centering
{\includegraphics[width = 8 cm]{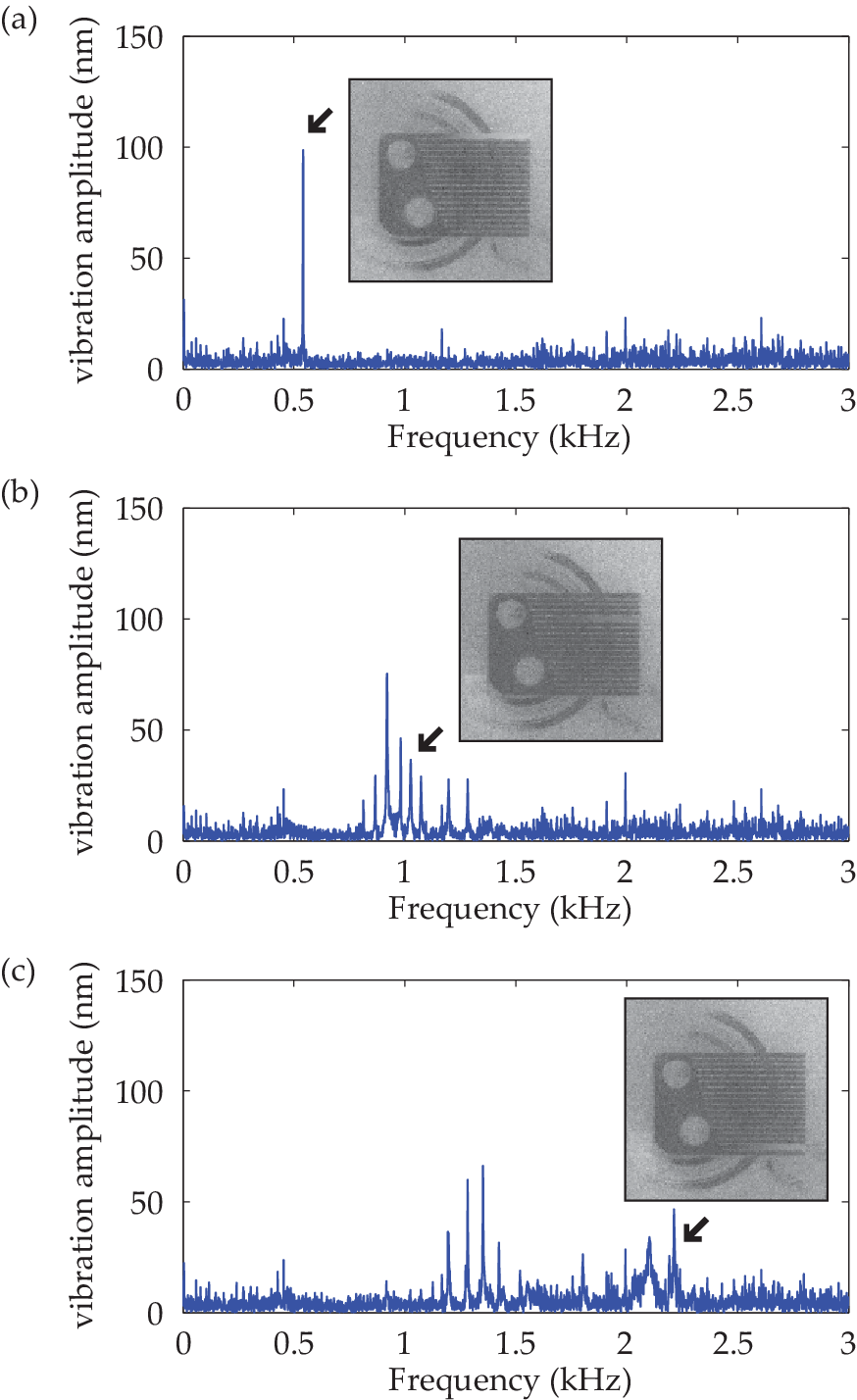}}
\caption{Vibration amplitude $z_0$ versus excitation frequency $\omega/(2\pi)$, averaged over the 1st (a), 5th (b), and 17th cantilevers. Insets: vibration amplitude maps at 541 Hz (a), 1006 Hz (b), and 2211 Hz (c).}
\label{fig_CantileversVibrationSpectra}
\end{figure}

To perform the measurement, an optical LO consisting of the combination of three frequency components shifted by $\Delta \omega _1$, $\Delta \omega _2$, and $\Delta \omega _3$ (Eq.~\ref{eq_E_LO}) was realized. Those shifts were chosen to yield non-symmetrical (non-opposed) algebraic modulation frequencies $\delta \omega _1$, $\delta \omega _2$, and $\delta \omega _3$ in the detector bandwidth in order to avoid spurious aliasing or cross-talk effects. To do so, we made use of phase-locked RF signals generated by frequency-synthesizers at the carrier frequency $\omega_{\rm C}$, $\omega_{\rm C} + \delta \omega_1'$, $\omega + \delta \omega_2'$, and $\omega_{\rm C} + \delta \omega_3'$, where the shifts used to generate the three RF bands were $\delta \omega_1' /(2\pi)= -1.25\, \rm Hz$, $\delta \omega_2' /(2\pi)= 6.25\, \rm Hz$, $\delta \omega_3' /(2\pi)= -2.5\, \rm Hz$ (see Tab.~\ref{tab_frequencies}). As sketched in Fig.~\ref{fig_setup}, those signals were mixed and combined to drive a set of acousto-optic modulators (AA Opto Electronic) designed to operate around $\omega_{\rm C}/(2\pi) = 80 \, \rm MHz$, from which opposite diffraction orders were selected. The resulting modulation frequencies of the holograms (Eq.~\ref{eq_Hm}) were $\delta \omega_1 = \delta \omega_1' + \delta \omega_2'$, $\delta \omega_2 = \delta \omega_3'$, $\delta \omega_3 = \delta \omega_1' - \delta \omega_2'$. The three sideband components  $\tilde{H}_{-1}$, $\tilde{H}_0$, and $\tilde{H}_1$, encoded at 5 Hz, -2.5 Hz, and -7.5 Hz respectively, are visible on the magnitude of the discrete Fourier spectrum (Eq.~\ref{eq_Hm}) reported in Fig.~\ref{fig_CCDSpectrum}. Spectral maps of the vibration amplitude $z_0(\omega)$ were calculated from a set of sequential measurements at each frequency. We took 3000 sequences of $N=8$ raw interferograms $\cal I$ sampled at a frame rate of  $\omega_{\rm S} / (2\pi) = 20 \, \rm Hz$, for excitation frequencies $\omega/(2 \pi)$ ranging from 0 Hz to 3 kHz, in 1 Hz steps (supply voltage of the piezo electric actuator : 10 mV). The first resonance frequencies of the lamellophone's cantilevers are observed between 500 Hz and 2500 Hz (Media 1). Plots of the vibration amplitude of the 1st, 5th, and 17th cantilevers versus the excitation frequency $\omega/(2\pi)$ are reported in  Fig.~\ref{fig_CantileversVibrationSpectra}. In particular, individual resonances of the 1st, 5th, and 17th cantilevers were identified around 541 Hz (Fig.~\ref{fig_CantileversVibrationSpectra}(a), inset), 1006 Hz  (Fig.~\ref{fig_CantileversVibrationSpectra}(b), inset), and 2211 Hz  (Fig.~\ref{fig_CantileversVibrationSpectra}(c), inset).\\

\begin{table}[b]
  \centering
  \begin{tabular}{cccc} \\ \hline
    \textbf{angular frequency} \qquad & \textbf{ value} \\ \hline
    $\delta \omega_1'$ & $- 1.25 \, \rm Hz$ \\
    $\delta \omega_2'$ & $6.25 \, \rm Hz$ \\
    $\delta \omega_3'$ & $- 2.5 \, \rm Hz$ \\ \hline
    $\delta \omega_1 = \delta \omega_1' + \delta \omega_2'$ & $5 \, \rm Hz$  \\
    $\delta \omega_2 = \delta \omega_3'$ & $-2.5 \, \rm Hz$  \\
    $\delta \omega_3 = \delta \omega_1' - \delta \omega_2'$  & $-7.5 \, \rm Hz$ \\ \hline
    $\Delta \omega_1 = -\omega - \delta \omega_1$  &  \\
    $\Delta \omega_2 =\delta \omega_2$ & \\
    $\Delta \omega_3 = \omega - \delta \omega_3$ &
    \\ \hline
  \end{tabular}
  \caption{Frequency shifts used in the experiments.}
\label{tab_frequencies}
\end{table}

\begin{figure}[]
\centering
{\includegraphics[width = 8 cm]{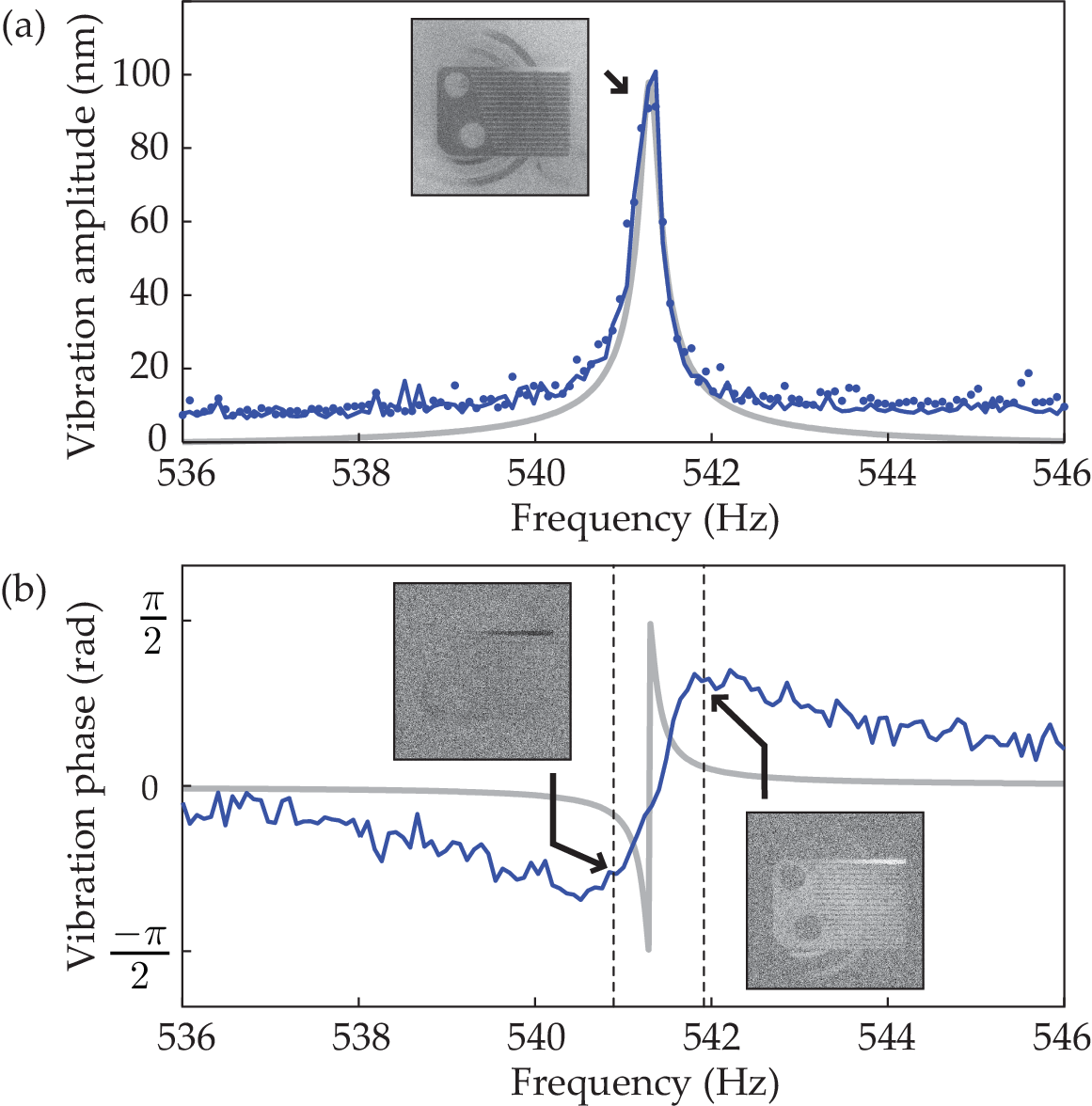}}
\caption{Vibration amplitude $z_0$ (a) and phase $\psi$ (b) averaged over the first cantilever, arrowed in Fig.~\ref{fig_AmplitudeAndPhaseImages}(b), versus excitation frequency $\omega/(2\pi)$. Insets : retrieved vibration amplitude and phase maps in the neighborhood of the resonance, reported in Fig.~\ref{fig_AmplitudeAndPhaseImages}(b-f). The points were obtained from a sequential measurement, the lines were obtained with the linear chirp (Eq.~\ref{eq_omega_M_Chirped}). The theoretical resonance lines in gray were determined from Eq.~\ref{eq_z0} and Eq.~\ref{eq_psi}. }
\label{fig_ChirpedSpectra}
\end{figure}

A finer screening around the resonance of the first cantilever was performed, from 248 sequences of $N = 8$ raw interferograms, for excitation frequencies ranging from 536 Hz to 546 Hz. The values of $z_0(\omega)$, spatially averaged over the extent of the resonant cantilever, arrowed in Fig.~\ref{fig_AmplitudeAndPhaseImages}(b),  are marked as points in the plot reported in Fig.~\ref{fig_ChirpedSpectra}(a). But $\psi_0(t)$ is randomized from the measurement of one frequency point to the next. Hence the phase retardation $\psi$ cannot be retrieved. To prevent reference phase randomization between measurements, a chirp of the frequency $\omega$ was performed. It consisted of a linear drift with time $t$ of the excitation frequency from $\omega_{\rm I} /(2 \pi)= 536 \, {\rm Hz}$ to $\omega_{\rm F}/(2 \pi) = 546 \, {\rm Hz}$ in $T = 99.2 \, {\rm s}$
\begin{equation}\label{eq_omega_M_Chirped}
\omega (t) = \omega_{\rm I} + (\omega_{\rm F} - \omega_{\rm I}) \, t / T
\end{equation}
Two out of the three components of the optical LO (Eq.~\ref{eq_E_LO}) were chirped as well via $\omega(t)$ throughout the acquisition. During the time lapse $T$, a sequence of $248 \times 8 = 1984$ raw interferograms $\cal I$ was steadily recorded. In that manner, $\psi_0$ remained constant. Amplitude $z_0(\omega)$ and phase $\psi(\omega)$ maps at each frequency $\omega$ were calculated from the demodulation of $N=8$ consecutive raw frames, reported in media 2. The complex values of these maps were averaged over the spatial extent of the first cantilever and reported in Fig.~\ref{fig_ChirpedSpectra} (blue lines). In agreement with Eq.~\ref{eq_psi}, for non-zero viscosity, a phase shift of $\sim\pi$ is observed at the resonance frequency $\omega_{\rm r} \sim 541.3 \, \rm Hz$. The theoretical resonance lines in Fig.~\ref{fig_ChirpedSpectra} were determined for a resonance frequency of the undamped oscillator $\omega_0/(2\pi)=541.3 \, \rm Hz$ and a dimensionless damping ratio $\zeta = 2\times10^{-5}$, for which Eq.~\ref{eq_z0} and Eq.~\ref{eq_psi} best fit the experimentally-measured vibration amplitude.\\ 

In conclusion, we performed optical path length modulation imaging by time-averaged heterodyne holography in off-axis and frequency-shifting conditions. To do so, modulation sidebands were acquired simultaneously through coherent frequency-division multiplexing. Additionally, a linear chirp of the excitation and the detection frequencies allowed to map the local mechanical phase retardation of a resonant cantilever with respect to the excitation signal without stroboscopy. The proposed method enabled robust and quantitative narrowband imaging of out-of plane vibration amplitude and phase.\\

We gratefully acknowledge support from Fondation Pierre-Gilles de Gennes (FPGG014), Agence Nationale de la Recherche (ANR-09-JCJC-0113, ANR-11-EMMA-046), r\'egion \^Ile-de-France (C'Nano, AIMA), and the "investments for the future" program (LabEx WIFI: ANR-10-LABX-24, ANR-10-IDEX-0001-02 PSL*).


\begin{thebibliography}{0}
\expandafter\ifx\csname natexlab\endcsname\relax\def\natexlab#1{#1}\fi
\expandafter\ifx\csname bibnamefont\endcsname\relax
  \def\bibnamefont#1{#1}\fi
\expandafter\ifx\csname bibfnamefont\endcsname\relax
  \def\bibfnamefont#1{#1}\fi
\expandafter\ifx\csname citenamefont\endcsname\relax
  \def\citenamefont#1{#1}\fi
\expandafter\ifx\csname url\endcsname\relax
  \def\url#1{\texttt{#1}}\fi
\expandafter\ifx\csname urlprefix\endcsname\relax\def\urlprefix{URL }\fi
\providecommand{\bibinfo}[2]{#2}
\providecommand{\eprint}[2][]{\url{#2}}

\end{thebibliography}


\begin{thebibliography}{10}

\bibitem{Bosseboeuf2003}
A.~Bosseboeuf and S.~Petitgrand.
\newblock Characterization of the static and dynamic behaviour of m (o) ems by
  optical techniques: status and trends.
\newblock {\em Journal of micromechanics and microengineering}, 13:S23, 2003.

\bibitem{Rembe2006}
C.~Rembe and A.~Drabenstedt.
\newblock Laser-scanning confocal vibrometer microscope: Theory and
  experiments.
\newblock {\em Rev. Scient. Instr.}, 77(8):083702--083702, 2006.

\bibitem{Kokkonen2008}
K.~Kokkonen and M.~Kaivola.
\newblock Scanning heterodyne laser interferometer for phase-sensitive
  absolute-amplitude measurements of surface vibrations.
\newblock {\em Appl. Phys. Lett.}, 92(6):063502--063502, 2008.

\bibitem{BramhavarPouet2009}
S.~Bramhavar, B.~Pouet, and T.~W. Murray.
\newblock Superheterodyne detection of laser generated acoustic waves.
\newblock {\em Appl. Phys. Lett.}, 94(11):114102--114102, 2009.

\bibitem{WhitmanKorpel1969}
RL~Whitman and A~Korpel.
\newblock Probing of acoustic surface perturbations by coherent light.
\newblock {\em Appl. Opt.}, 8(8):1567--1576, 1969.

\bibitem{DeLaRue1972}
R.M. De~La~Rue, R.F. Humphryes, I.M. Mason, and E.A. Ash.
\newblock Acoustic surface wave amplitude and phase measurements using laser
  probes.
\newblock {\em Proc. IEE.}, 119(2):117--126, 1972.

\bibitem{Stegeman1976}
G.~Stegeman.
\newblock Optical probing of surface waves and surface wave devices.
\newblock {\em IEEE Transactions on Sonics Ultrasonics}, 23:33--63, 1976.

\bibitem{Monchalin1985}
Jean-Pierre Monchalin.
\newblock Heterodyne interferometric laser probe to measure continuous
  ultrasonic displacements.
\newblock {\em Rev. Scient. Instr.}, 56(4):543--546, 1985.

\bibitem{WagnerSpicer1987}
J.W. Wagner and J.B. Spicer.
\newblock Theoretical noise-limited sensitivity of classical interferometry.
\newblock {\em J. Opt. Soc. Am. B}, 4(8):1316--1326, 1987.

\bibitem{RoyerDieulesaint1989}
D~Royer and E~Dieulesaint.
\newblock Mesures optiques de d{\'e}placements d'amplitude 10\^{}-4 {\`a}
  10\^{} 2 angstr{\"o}m. application aux ondes {\'e}lastiques.
\newblock {\em Rev. Phys. Appl.}, 24(8):833--846, 1989.

\bibitem{RoyerDieulesaint1986}
D~Royer and E~Dieulesaint.
\newblock Optical probing of the mechanical impulse response of a transducer.
\newblock {\em Appl. Phys. Lett.}, 49(17):1056--1058, 1986.

\bibitem{JiaBoumiz1993}
X~Jia, A~Boumiz, and G~Quentin.
\newblock Laser interferometric detection of ultrasonic waves propagating
  inside a transparent solid.
\newblock {\em Appl. Phys. Lett.}, 63(16):2192--2194, 1993.

\bibitem{RoyerKmetik1992}
D.~Royer and V.~Kmetik.
\newblock Measurement of piezoelectric constants using an optical heterodyne
  interferometer.
\newblock {\em Electronics Letters}, 28(19):1828--1830, 1992.

\bibitem{Kimachi2010}
Akira Kimachi.
\newblock Real-time heterodyne speckle pattern interferometry using the
  correlation image sensor.
\newblock {\em Applied optics}, 49(35):6808--6815, 2010.

\bibitem{PatelAchamfuoYeboah2011}
Rikesh Patel, Samuel Achamfuo-Yeboah, Roger Light, and Matt Clark.
\newblock Widefield heterodyne interferometry using a custom cmos modulated
  light camera.
\newblock {\em Optics Express}, 19(24):24546--24556, 2011.

\bibitem{Powell1965}
R.~L. Powell and K.~A. Stetson.
\newblock Interferometric vibration analysis by wavefront reconstruction.
\newblock {\em J. Opt. Soc. Am.}, 55:1593, 1965.

\bibitem{Aleksoff1969}
C.~C. Aleksoff.
\newblock Time average holography extended.
\newblock {\em Appl. Phys. Lett}, 14:23, 1969.

\bibitem{Stetson1970}
Karl~A. Stetson.
\newblock Effects of beam modulation on fringe loci and localization in
  time-average hologram interferometry.
\newblock {\em J. Opt. Soc. Am.}, 60(10):1378, 1970.

\bibitem{Levitt1976}
JA~Levitt and KA~Stetson.
\newblock Mechanical vibrations: mapping their phase with hologram
  interferometry.
\newblock {\em Applied Optics}, 15(1):195--199, 1976.

\bibitem{UedaMiida1976}
M.~Ueda, S.~Miida, and T.~Sato.
\newblock Signal-to-noise ratio and smallest detectable vibration amplitude in
  frequency-translated holography: an analysis.
\newblock {\em Appl. Opt.}, 15(11):2690--2694, 1976.

\bibitem{PicartLeval2003}
P.~Picart, J.~Leval, D.~Mounier, and S.~Gougeon.
\newblock Time-averaged digital holography.
\newblock {\em Opt. Lett.}, 28:1900--1902, 2003.

\bibitem{PsotaLedl2012}
P.~Psota, V.~Ledl, R.~Dolecek, J.~Erhart, and V.~Kopecky.
\newblock Measurement of piezoelectric transformer vibrations by digital
  holography.
\newblock {\em IEEE Transactions on Ultrasonics, Ferroelectrics and Frequency
  Control}, 59(9):1962--1968, 2012.

\bibitem{VerrierAtlan2013}
Nicolas Verrier and Michael Atlan.
\newblock Absolute measurement of small-amplitude vibrations by time-averaged
  heterodyne holography with a dual local oscillator.
\newblock {\em Opt. Lett.}, 38(5):739--741, 2013.

\bibitem{Pedrini2006}
Giancarlo Pedrini, Wolfgang Osten, and Mikhail~E. Gusev.
\newblock High-speed digital holographic interferometry for vibration
  measurement.
\newblock {\em Appl. Opt.}, 45(15):3456--3462, May 2006.

\bibitem{PerezLopez2006}
C.~Perez-Lopez, M.H. De~la Torre-Ibarra, and F.~Mendoza~Santoyo.
\newblock Very high speed cw digital holographic interferometry.
\newblock {\em Optics Express}, 14(21):9709--9715, 2006.

\bibitem{Lokberg1976}
Ole~J. L{\o}kberg and K{\aa}re H{\o}gmoen.
\newblock Vibration phase mapping using electronic speckle pattern
  interferometry.
\newblock {\em Appl. Opt.}, 15(11):2701--2704, Nov 1976.

\bibitem{Petitgrand2001}
S.~Petitgrand, R.~Yahiaoui, K.~Danaie, A.~Bosseboeuf, and JP~Gilles.
\newblock 3d measurement of micromechanical devices vibration mode shapes with
  a stroboscopic interferometric microscope.
\newblock {\em Optics and lasers in engineering}, 36(2):77--101, 2001.

\bibitem{LevalPicart2005}
Julien Leval, Pascal Picart, Jean~Pierre Boileau, and Jean~Claude Pascal.
\newblock Full-field vibrometry with digital fresnel holography.
\newblock {\em Appl. Opt.}, 44(27):5763--5772, Sep 2005.

\bibitem{VerrierGrossAtlan2013}
Nicolas Verrier, Michel Gross, and Michael Atlan.
\newblock Phase-resolved heterodyne holographic vibrometry with a strobe local
  oscillator.
\newblock {\em Optics letters}, 38(3):377--379, 2013.

\bibitem{AtlanGross2007JOSAA}
M.~Atlan and M.~Gross.
\newblock Spatiotemporal heterodyne detection.
\newblock {\em Journal of the Optical Society of America A}, 24(9):2701--2709,
  2007.

\bibitem{PaturzoMemmolo2009}
M.~Paturzo, P.~Memmolo, A.~Tulino, A.~Finizio, and P.~Ferraro.
\newblock Investigation of angular multiplexing and de-multiplexing of digital
  holograms recorded in microscope configuration.
\newblock {\em Opt. Express}, 17(11):8709--8718, 2009.

\bibitem{Kiire2012}
Tomohiro Kiire, Daisuke Barada, Jun ichiro Sugisaka, Yoshio Hayasaki, and
  Toyohiko Yatagai.
\newblock Color digital holography using a single monochromatic imaging sensor.
\newblock {\em Opt. Lett.}, 37(15):3153--3155, Aug 2012.

\bibitem{Tahara2012}
Tatsuki Tahara, Akifumi Maeda, Yasuhiro Awatsuji, Takashi Kakue, Peng Xia,
  Kenzo Nishio, Shogo Ura, Toshihiro Kubota, and Osamu Matoba.
\newblock Single-shot dual-illumination phase unwrapping using a single
  wavelength.
\newblock {\em Opt. Lett.}, 37(19):4002--4004, Oct 2012.

\bibitem{Cuche2000}
Etienne Cuche, Pierre Marquet, and Christian Depeursinge.
\newblock Spatial filtering for zero-order and twin-image elimination in
  digital off-axis holography.
\newblock {\em Applied Optics}, 39(23):4070, 2000.

\end{thebibliography}

\end{document}